\documentclass[apsrev,prb,twocolumn]{revtex4-2}
\usepackage{amssymb,amsfonts,amsmath}
\usepackage{adjustbox}
\usepackage{hyperref}
\usepackage{color}
\usepackage{xr-hyper}
\usepackage[markup]{changes}
\usepackage{textgreek}
\usepackage{float}

\begin{document}

\title{Thermoelectricity of Tin Selenide Monolayers Across a Structural Phase Transition}
\author{John W. Villanova}
\email{jvillano@uark.edu}
\affiliation{Department of Physics, University of Arkansas, Fayetteville, Arkansas 72701, United States}
\author{Salvador Barraza-Lopez}
\email{sbarraza@uark.edu}
\affiliation{Department of Physics, University of Arkansas, Fayetteville, Arkansas 72701, United States}

\begin{abstract}
SnSe monolayers experience a temperature induced two-dimensional Pnm2$_1 \to$ P4/nmm structural transformation precipitated by the softening of vibrational modes. The standard theoretical treatment of thermoelectricity---which relies on a zero temperature phonon dispersion and on a zero temperature electronic structure---is incapable of describing thermoelectric phenomena induced by structural transformations. Relying on structural data obtained from {\em ab initio} molecular dynamics calculations that is utilized in a non-standard way to inform of electronic and vibrational transport coefficients, the present work establishes a general route to understand thermoelectricity across phase transitions. Similar to recent experimental observations pointing to an overestimated thermoelectric figure of merit $ZT$ past the transition temperature, our work indicates a smaller $ZT$ when compared to its value predicted by the standard paradigm. Its decrease is related to the dramatic changes in the electrical conductivity and lattice thermal conductivity as the structural transformation ensues. Though exemplified on a SnSe monolayer, the method does not have any built-in assumptions concerning dimensionality, and thus applicable to arbitrary thermoelectric materials in one, two, and three dimensions.
\end{abstract}

\maketitle
\date{today}

\section{Introduction}
According to the U.S. National Academies, energy supply is one of the greatest challenges of the century, and energy diversification is needed.\cite{NAP12619} Thermoelectric materials are candidates for such diversification. Nevertheless and despite its longevity, thermoelectricity theory has clear mandates for improvement. For instance, Mingo and coworkers indicate in their Perspective Review of {\em ab initio} phonon thermal transport that higher order phonon scattering is desirable within thermoelectric frameworks, and that such scattering could in principle be obtained from {\em ab initio} molecular dynamics.\cite{mingo} Furthermore, another recent review on thermoelectricity by Snyder and coworkers clearly states that: ``our models and current understanding break down as conditions deviate from an idealized, static system, {\em for example, at temperatures above absolute zero},'' and that  ``finding ways to {\em move beyond our current reliance on the ground state electronic and phonon band structures} ... will be key to future progress in this area''\cite{review2018} (emphasis ours). This work fills that void by establishing a formalism whereby finite-temperature electronic and vibrational behavior inform thermoelectric properties. The methodology introduced in the present work was inspired by and directly responds to vocal needs recently expressed from leaders within the thermoelectric community. Given that this new methodology has immediate consequences for the design of thermoelectric materials, it is of interest to the broader materials community.

As a critical example underlying the importance of the present results, bulk SnSe has been argued to display an extremely large thermoelectric figure of merit $ZT$ at high temperature near to its Pnma $\to$ Cmcm structural transformation,\cite{ZHAO14,fullydense,CCHANG18} and multiple theoretical works based on the preponderant {\em ab initio} methodologies\cite{boltztrap,boltztrap2,shengbte} have supported and even provided the rationale for these experimental claims (see {\em e.g.}, \cite{DING15,PhysRevB.92.115202,scirep2016} among multiple others). Nevertheless, the existence of such a high $ZT$ has been put in doubt in recent detailed experimental work.\cite{CHEN18,AGNE19} In a nutshell, the issue at hand is a serious underestimation of the lattice thermal conductivity at the onset of the structural transition in previous (experimental {\em and} theoretical) work.

Grown on a graphitic substrate, single layers of SnSe have just been experimentally realized.\cite{arxivKai} And while bulk SnSe undergoes a phase transition at temperatures as high as 900 K,\cite{SnSebulk1} these SnSe monolayers on graphite display a lower critical temperature, closer to 400 K.\cite{arxivKai} Due to the high expense of performing {\em ab initio} molecular dynamics of SnSe on graphite on samples of experimentally-relevant sizes, most theoretical work on these monolayers is carried out on freestanding samples. SnSe monolayers (as a prototypical example of isostructural group-IV monochalcogenide monolayers) are intriguing materials on their own:\cite{RMP} They are structurally-stable binary semiconductors with an intrinsic in-plane electric dipole moment\cite{fei_apl_2015_ges_gese_sns_snse,KAICHANG} in their ground-state Pnm2$_1$ structural configuration.\cite{FWANG15,rodin_prb_2016_sns} Additional studies concerning electronic valleys,\cite{rodin_prb_2016_sns,slawinska_2d_mat_2018_mmls,KaiPRL} optical properties,\cite{b2,b4,Wangeaav9743} and even thermoelectric properties\cite{SUN19,DING2019} have been reported, {\em all} in reference to ground state atomistic structures with a Pnm2$_1$ group symmetry.

An incipient study by Wang and coworkers \cite{FWANG15} of the thermoelectric properties of freestanding SnSe monolayers contains the following standard elements from {\em ab initio} thermoelectricity theory: they found electronic transport coefficients by solving for the semiclassical Boltzmann equation within the relaxation time approximation, as implemented in the BoltzTraP program.\cite{boltztrap,boltztrap2} Lattice thermal conductivity and phonon lifetimes were obtained using Mingo and coworkers' ShengBTE code.\cite{shengbte} Their calculation of vibrational properties included up to third-order interatomic force constants.Furthermore, and consistent with standard methods,\cite{mingo,review2018} the unit cell considered for the calculation of electronic and vibrational transport coefficients possessed a Pm2$_1$n symmetry (or Pnm2$_1$ with our choice of principal axes), {\em and it is determined at zero temperature}. These calculations were carried out with the PBE exchange-correlation functional.\cite{PBE} They acknowledge the following points concerning the structural transition of SnSe: (a) {\em bulk} SnSe undergoes a phase transition at around 750-800 K ($T_c$ is actually higher\cite{SnSebulk1}); (b) the single-layered SnSe sheet may also experience a phase transition upon heating; (c) ``the phase transition is beyond the scope of our work;'' and (d) thermoelectric properties are determined at the (``medium-high'') 300-700 K range.\cite{FWANG15} Two crucial questions left unanswered in their work are the magnitude of the critical temperature $T_c$ in a SnSe monolayer, and the effect of the structural transition on thermoelectric properties. Clearly, answering these questions has consequences for the theoretical treatment of {\em arbitrary} thermoelectric materials undergoing a phase transition.

After the work by Wang {\em et al.},\cite{FWANG15} we have developed a detailed understanding of the structural transition undergone by SnSe monolayers.\cite{MEHB16_nano,MEHB16_prl,other4,SHIVA19,newarXiv,RMP} These are its main points: (a) Distances among second and third nearest neighbors turn equal ($d_2=d_3$) for temperatures above $T_c$ ($d_1$ and $d_2$ were reported, but $d_3$ was not). (b) The critical temperature $T_c$ depends on the ensemble being utilized (NVT---in which the volume remains fixed---{\em versus} NPT---in which pressure remains fixed): $T_c$ is larger when using the NVT ensemble. Additionally, the phases above $T_c$ are Pnmm when using the NVT ensemble, and P4/nmm when the NPT ensemple is employed.\cite{newarXiv} (c) $T_c$ also depends on the exchange-correlation functional employed.\cite{SHIVA19,newarXiv}

Taking the choices from Wang and coworkers (an NVT ensemble and the PBE exchange-correlation functional), a SnSe monolayer transitions from the low-temperature Pnm2$_1$ phase onto a Pnmm phase at $T_c=320$ K,\cite{newarXiv} which lies right at the lower end in the temperature range they considered.\cite{FWANG15} Since they did not determine $T_c$, no mention of the Pnmm phase can be found in their work. This fact shows that their thermoelectric coefficients at all temperatures were obtained using standard methods, which require the unit cell to be the zero-temperature one with a Pnm2$_1$ symmetry, as they themselves state. In short, their study stayed short of linking finite temperature atomistic and electronic structure to thermoelectric properties.

Further, structural transformations are driven by the softening optical phonon modes,\cite{newarXiv} which can be hardly accounted for with the standard perturbative approach. This fact is acknowledged by the creators of state-of-the-art tools for the {\em ab initio} treatment of thermoelectricity themselves.\cite{mingo}. Any approach  to vibrational properties that is not perturbative---like the one to be laid out here---is not standard and is sought after within thermoelectricity theory.\cite{mingo}

The thermoelectric figure of merit $ZT$ is defined as:\cite{boltztrap,review2018}
\begin{equation}\label{eq:eq1}
ZT=\frac{\sigma S^2T}{\kappa_{e}+\kappa_{l}},
\end{equation}
where $\sigma$ is the electrical conductivity, $S$ is the Seebeck coefficient, $T$ is the temperature, $\kappa_e$ is the electronic contribution to the thermal conductivity, and $\kappa_l$ is the lattice thermal conductivity.
We calculate the thermoelectric figure of merit $ZT$ across the 2D structural phase transition, something not yet done by theoretical calculations which had previously utilized the zero temperature structure in a temperature range which could not capture the transformation. The factors entering into $ZT$ are calculated with reference to structural data at finite temperature, representing a crucial departure from the common reliance on ground state electronic and phonon band structures. With this new formalism, one observes a sudden increase in the electronic conductivity and in the electronic and lattice contributions to the thermal conductivity, happening at temperatures driving the structural transformation.\cite{other4,newarXiv} The theoretical examination of the lattice thermal conductivity accounts for phonon anharmonicity non-perturbatively throughout the structural transformation. $ZT$ decreases substantially near the onset of the structural transformation, suggesting that the standard methods for calculating thermoelectric properties have and will overestimate $ZT$ when they ignore structural transformations in systems with significant anharmonicity. These theoretical results are in agreement with recent experimental results questioning record high values of $ZT$ on materials undergoing structural phase transitions.\cite{CHEN18,AGNE19}
Though exemplified on a 2D ferroelectric, the process applies to any material undergoing solid-to-solid structural transformations and may be applied to 1D and 3D materials as well.

\section{Results and discussion}

\subsection{Thermally-driven structural transformation of a SnSe monolayer}

We use finite temperature structural data obtained from molecular dynamics calculations to determine all of the factors contributing to $ZT$. These structural transformations have been studied by means of {\em ab initio} molecular dynamics calculations within the isobaric-isothermal (NPT) ensemble,\cite{other4,newarXiv} and the data in this study was obtained on a 16$\times$16 SnSe monolayer supercell containing 1024 atoms with a 1.5 fs time resolution for about 28,000 fs, and for more than ten temperatures (in contrast to previous molecular dynamics calculations using 64 atoms, and running for 10,000 fs at five different temperatures\cite{FWANG15}; see Methods).

As previously indicated,\cite{other4,newarXiv} the magnitude of $T_c$ depends on structural constraints. Imposing a constant volume is inconsistent with previous results on these transformations,\cite{KAICHANG} which support a transformation with a change in area.\cite{other4} The NPT ensemble is, on the other hand, consistent with a transition with a change in area and has been chosen in this study for that reason. Furthermore, the dependency of thermoelectric coefficients on the exchange-correlation functional has been acknowledged,\cite{mingo} and the lone pairs on SnSe monolayers call for a treatment of the electronic structure beyond PBE; a detailed discussion has been given elsewhere.\cite{SHIVA19} The vdW-DF-cx van der Waals exchange-correlation functional is employed here for that reason.\cite{soler,BH}

\begin{figure}[tb]
\begin{center}
\includegraphics[width=0.48\textwidth]{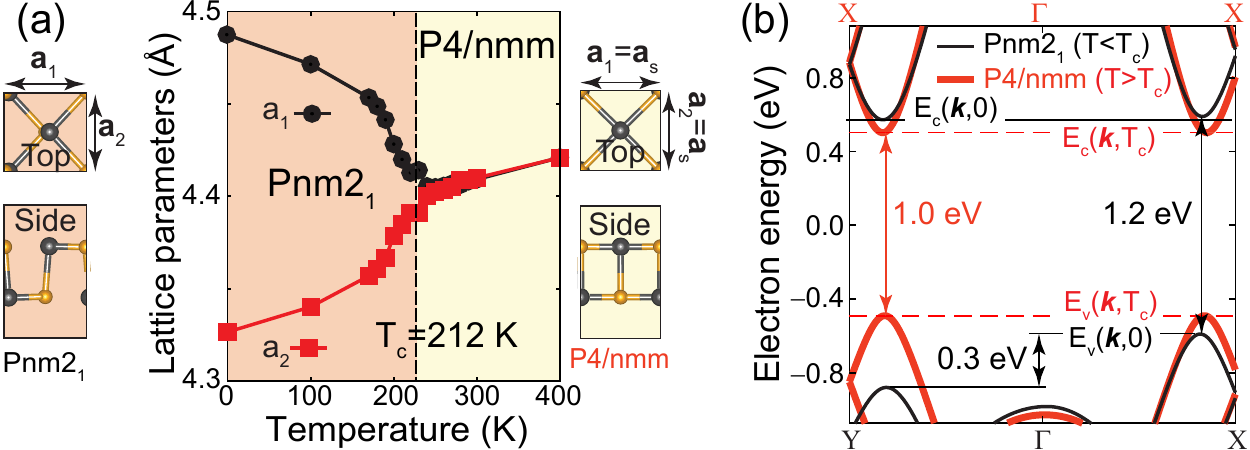}
\caption{(a) The temperature evolution of in-plane lattice parameters $a_1$ and $a_2$ for a SnSe monolayer demonstrates a structural transformation from an orthorhombic unit cell below $T_c$ into a unit cell with tetragonal symmetry above $T_c=212$ K.\cite{other4,newarXiv} Top and side views of the atomistic arrangements of these two structural phases are shown. (b) Evolution of the valence ($v$) and conduction ($c$) bands of the SnSe monolayer as a result of the Pnm2$_1 \to$ P4/nmm structural transformation.}\label{fig:fig1}
\end{center}
\end{figure}

Figure \ref{fig:fig1}(a) displays the average lattice parameters $a_1$ and $a_2$ of a freestanding SnSe monolayer as a function of temperature. A fit of critical exponents yields $T_c=212$ K.\cite{other4} At temperatures equal or above $T_c$, the SnSe monolayer turns paraelectric, which implies that the unit cell develops two mirror planes and a four-fold symmetry consistent with a P4/nmm phase above $T_c$. As indicated before, $T_c \sim$ 400 K in experiments of SnSe monolayers on a graphitic substrate.\cite{arxivKai} The higher $T_c$ observed in experiment is attributed to the interaction of the SnSe monolayer with its supporting substrate. In any event, Figure\ref{fig:fig1}(a), experiment, and the fact that the standard thermoelectric theory relies on zero-temperature structural data all demonstrate that an incorrect atomistic symmetry  (Pnm2$_1$) has been previously employed to determine the thermoelectric properties of a SnSe monolayer within the reported 300-700 K temperature range.\cite{FWANG15,SUN19,DING2019} The improper use of the zero-temperature atomistic structure should lead to striking discrepancies with experiment. The signature of the structural transition is the confluence of the in-plane lattice parameters at a critical temperature $T_c=212$ K that is determined from a critical exponent fit.\cite{other4} The appropriate and consistent use of finite-temperature information to determine the full thermoelectric properties is delineated next. No such comprehensive and novel development exists within the literature.

\subsection{Thermal evolution of the electronic structure}

As seen in Figure \ref{fig:fig1}(b), structural changes modify the electronic structure. The band structure shown in black corresponds to the Pnm2$_1$ atomistic structure at zero temperature; {\em i.e.,} to the electronic structure employed at all temperatures (300 to 700 K) in works that rely on the standard formalism.\cite{FWANG15,SUN19,DING2019} Those works use a band structure that no longer corresponds with the atomistic structure above 212 K on a freestanding SnSe monolayer,\cite{other4} or above 400 K for a SnSe monolayer on a graphitic substrate.\cite{arxivKai} The electronic band structure shown in red in Figure \ref{fig:fig1}(b) corresponds to the average atomistic structure of the freestanding SnSe monolayer above 212 K. There, the two hole valleys turn degenerate due to the enhanced tetragonal symmetry of the P4/nmm symmetry group. The continuous structural transformation results in a continuous change in the electronic band structure, which in Figure \ref{fig:fig1}(b), smoothly evolves with temperature from the one shown in black at $T=0$ K into the one shown in red at temperatures larger than $T_c$. In what follows, the electronic band structure is labeled $E_{\alpha}(\mathbf{k},T)$, where $\alpha$ is the band index ($\alpha=v$ for valence and $c$ for conduction band, respectively), in order to emphasize its direct dependence in temperature through the use of the average unit cell in our molecular dynamics calculations at finite temperature. It is important to note that we abandon the rigid band approximation by using the proper temperature dependent electronic structure rather than relying on the zero temperature band structure at this stage.

\begin{figure}[tb]
\begin{center}
\includegraphics[width=0.48\textwidth]{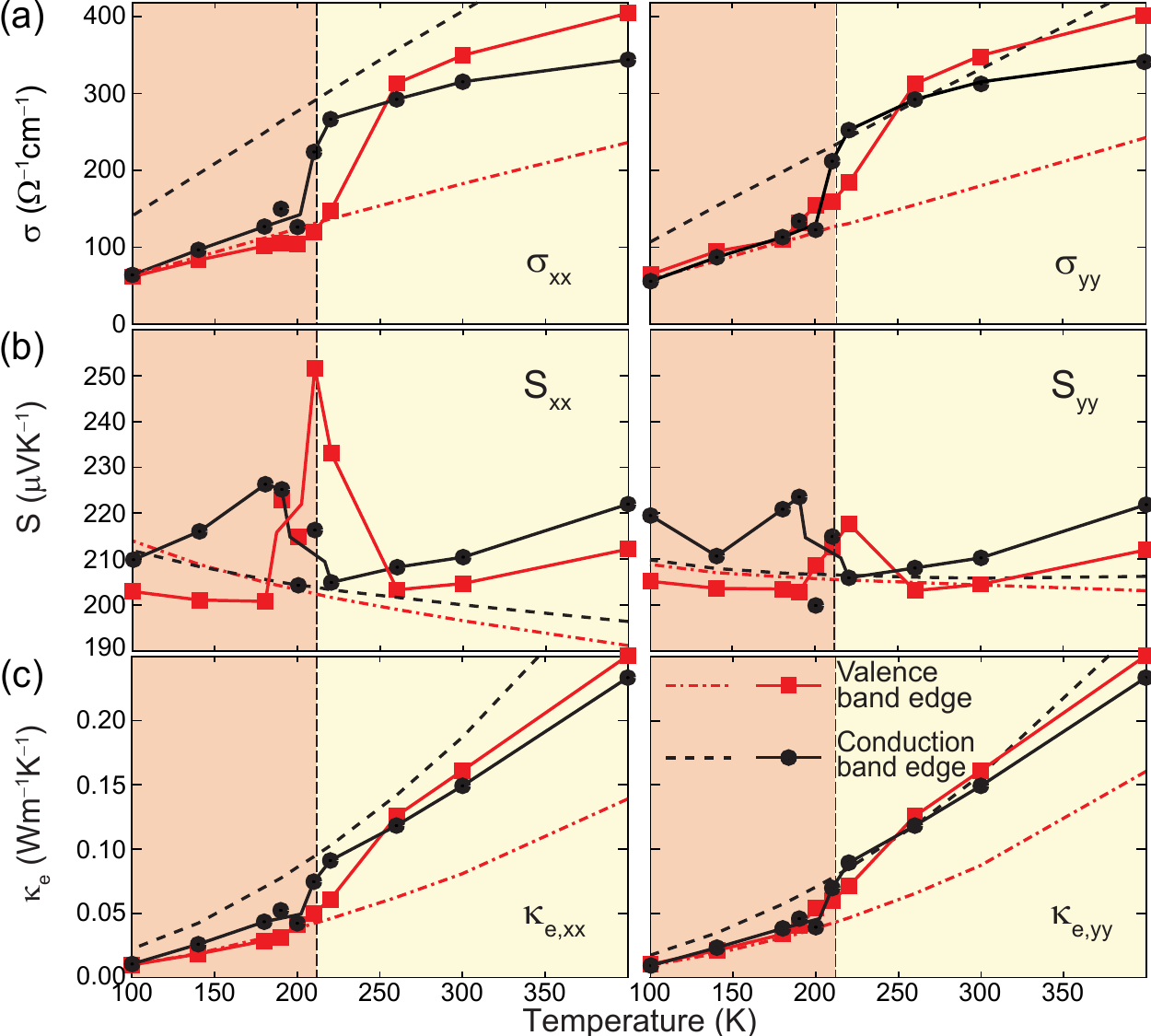}
\caption{(a) Electrical conductivity $\sigma$, (b) Seebeck coefficient $S$, and (c) electronic contribution to thermal conductivity $\kappa_e$ along the $\mathbf{a}_1$ ($x$) and $\mathbf{a}_2$ ($y$) directions versus $T$, respectively, for $\mu$ set at the conduction and valence band edges. Note the upticks at $T_c$ due to a temperature-induced band alignment and the enhanced symmetry past $T_c$, both absent features in the standard approach (dashed and dash-dotted curves).}\label{fig:fig2}
\end{center}
\end{figure}

\subsection{Electronic contributors to the thermoelectric figure of merit}

Knowing the effect of temperature on lattice parameters and on the electronic structure $E_{\alpha}(\mathbf{k},T)$, one can calculate the electrical conductivity $\sigma_{ij}$ tensor:\cite{review2018}
\begin{widetext}
\begin{equation}\label{eq:eq2}
\sigma_{ij}(T)= \frac{e^2}{\Omega(T)}\sum_{\mathbf{k},\alpha} \mathrm{v}_{\alpha,i}(\mathbf{k},T) \mathrm{v}_{\alpha,j}(\mathbf{k},T) \tau_e(T) \left(-\frac{\partial f(E-\mu,T)}{\partial E}\right),
\end{equation}
\end{widetext}
where $i=x,y$ and $j=x,y$ represent cartesian coordinates, $e$ is the electron charge, and $\hbar$ is the reduced Planck constant. $\Omega(T)$ is the temperature dependent volume of the unit cell, $\tau_e$ is the electron relaxation time, $\mathrm{v}_{\alpha,i}(\mathbf{k},T)=\frac{1}{\hbar}\frac{\partial E_{\alpha}(\mathbf{k},T)}{\partial k_i}$ is the band group velocity, and $f(E-\mu,T)$ is the Fermi-Dirac distribution at chemical potential $\mu$. The temperature dependence of $\sigma$ in the standard {\em ab initio} approach to thermoelectricity only enters through $\tau$ and $f(E-\mu,T)$. Here, on the other hand, electron group velocities and the unit cell area are made temperature-dependent by the use of finite-temperature structural data. Although the electron relaxation time $\tau_e$ may also be temperature dependent, a temperature-independent magnitude of $10^{-14}$ is assigned (comparable to $h/(1 \text{eV})$ for a 1 eV band width with $h$ the Planck constant) in accordance within previous estimates.\cite{FWANG15,HU17,DING15,GONZ17,DING2019} Equation \ref{eq:eq2} takes a sum over all bands, but the derivative of the Fermi-Dirac distribution selects contributions from bands $E_{\alpha}(\mathbf{k},T)$ having energies near the chemical potential $\mu$. Every Equation shown here takes appropriate care of dimensionality implicitly, and this general approach can be applied unchanged to arbitrary materials undergoing phase transitions, hence its importance.

Figure \ref{fig:fig2}(a) showcases in dashed (dashed-dotted) lines the electron (hole) conductivity with $\mu$ set exactly at the conduction (valence) band edge using a zero-temperature volume $\Omega$ and zero-temperature electronic structure to compare the prediction from this new method with the standard paradigm. As may be expected, the temperature dependent method, shown by solid lines and red squares (black circles) for $\mu$ set at the valence (conduction) band edge, tracks closely with the standard paradigm before the structural transformation. But there is a marked increase in the electrical conductivity at $T_c$ as the valence and conduction bands align with the increased symmetry above $T_c$, and it reflects the enhanced symmetry for temperatures above $T_c$. For $T>T_c$, the standard paradigm predicts greater electron conductivity as compared to hole conductivity, but the temperature dependent formalism indicates the opposite trend for $T\ge 250$ K.

The next term in the numerator of Equation \ref{eq:eq1} is the Seebeck coefficient $S$ which is obtained by dividing the following expression
\begin{widetext}
\begin{equation}\label{eq:eq5}
[\sigma S(T,\mu)]_{ij}= \frac{-e}{T\Omega(T)}\sum_{\mathbf{k},\alpha} \mathrm{v}_{\alpha,i}(\mathbf{k},T) \mathrm{v}_{\alpha,j}(\mathbf{k},T) \tau_e \left(-\frac{\partial f(E-\mu,T)}{\partial E}\right) (E_{\alpha}(\mathbf{k},T)-\mu)
\end{equation}
\end{widetext}
by Equation \ref{eq:eq2}. $S$, as depicted in Figure \ref{fig:fig2}(b), was computed with $\mu$ set at either the valence or conduction band edges. As was the case for $\sigma$, $S$ is asymmetric in the usual formalism above $T_c$ (after the system has actually acquired four-fold symmetry), failing to acknowledge the structural phase transition. $S$ also exhibits a peak at $T_c$ for $S_{xx}$ and a subdued peak for $S_{yy}$. When contrasted with the temperature evolution of $\sigma$, the Seebeck coefficient is otherwise roughly constant over the temperature range investigated, and the finite temperature estimates at over 200 \textmu V\,K$^{-1}$ are comparable to those obtained within the standard paradigm.

The denominator in Equation \ref{eq:eq1} contains the electrical and lattice contributions to the thermal conductivity, $\kappa_e$ and $\kappa_l$, respectively. The electronic contribution to the thermal conductivity $\kappa_{e,ij}$ has two terms,
\begin{equation}
\kappa_{e,ij}(T,\mu) = {\cal K}_{ij}(T,\mu)-T[\sigma S(T,\mu)]^2_{ij}\sigma(T,\mu)_{ij}^{-1},
\end{equation}
with the first contribution being
\begin{widetext}
\begin{equation}
{\cal K}_{ij}(T,\mu)= \frac{1}{T\Omega(T)}\sum_{\mathbf{k},\alpha} \mathrm{v}_{\alpha,i}(\mathbf{k},T) \mathrm{v}_{\alpha,j}(\mathbf{k},T) \tau_e \left(-\frac{\partial f(E-\mu,T)}{\partial E}\right) (E_{\alpha}(\mathbf{k},T)-\mu)^2,
\end{equation}
\end{widetext}
and the second term expressible from Equations \ref{eq:eq2} and \ref{eq:eq5}. $\kappa_{e}$ (Figure \ref{fig:fig2}(c)) displays a trend similar to the one observed for the electrical conductivity in Figure \ref{fig:fig2}(a). The most dramatic discrepancy between the present method and the previous one occurs for the valence band, given that there is a 0.3 eV shift to bring the hole pockets to the same energy beyond $T_c$ which is completely ignored by the usual rigid band approximation. In addition, the transport coefficients turn symmetric past $T_c$. The enhanced symmetry seen in all subplots of Figure \ref{fig:fig2} past $T_c$ is a signature that could be employed experimentally to determine whether this finite temperature approach to thermoelectricity surpasses the current state-of-the-art. The theoretical treatment that most closely resembles ours calls for a discontinuous jump of $\sigma$ at $T_c$, and does not contain a prescription for $\kappa_e$ across the structural transition.\cite{scirep2016} As indicated next, calculating the lattice thermal conductivity requires additional methods to collect the phonon frequencies and lifetimes from the molecular dynamics data.

\begin{figure*}[tb]
\begin{center}
\includegraphics[width=0.96\textwidth]{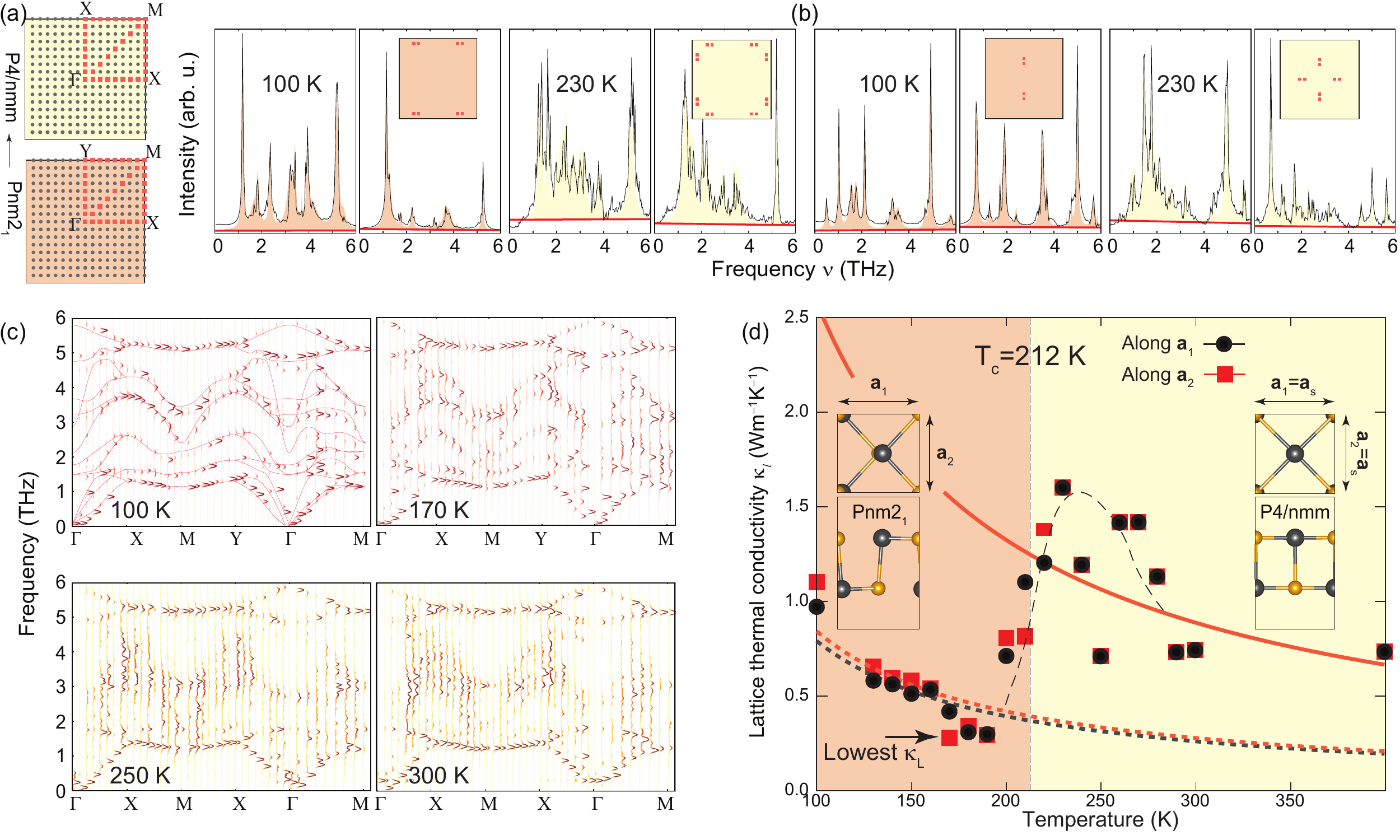}
\caption{(a) First Brillouin zone below and above the critical temperature $T_c=212$ K.\cite{other4,newarXiv} (b) Samples of power spectra at two consecutive $k-$points at temperatures below and above $T_c$. (c) Phonon spectrum at finite temperature below $T_c$ (upper two subplots) and above $T_c$ (lower two subplots) along the path shown in red in (a). Optical phonon modes in the range of 2 to 4 THz are seen to soften at $T>T_c$. Red curves are zero temperature phonons without anharmonic corrections and serve as a guide to the eye. (d) Circles and squares represent $\kappa_l$ as obtained from the phonon spectrum at finite $T$ along the $x-$ ($a_1$) and $y-$ ($a_2$) directions. The solid line is the fit to the data above $T=190$ K and the dotted black (red) line is the fit to the data below $T=190$ K along the $a_1$ ($a_2$) direction. These latter fits are indistinguishable from the temperature behavior derived from a zero temperature phonon spectrum, but underestimate $\kappa_l$ above $T_c$. The black dashed curve is a guide to the eye, showing the sudden increase in $\kappa_l$ at and beyond $T_c$.}\label{fig:fig3}
\end{center}
\end{figure*}

\subsection{Vibrational spectrum at finite temperature}


In materials at the onset of a structural transformation such as bulk SnSe,\cite{LI15,chemmat} SnTe monolayers,\cite{KAICHANG} and others, ``anharmonicity drives the crystal past the zero-temperature structure onto a {\em new crystalline phase} for which the zero-temperature electronic and zero-temperature phonon dispersions may no longer carry meaning.''\cite{Heremans2} Yet the present theory paradigm of thermoelectricity assumes that anharmonicity can be accounted for perturbatively, and it ignores structural transformations altogether.\cite{review2018,JAP2019,boltztrap,boltztrap2,shengbte} Said more explicitly, ``vibrational properties are typically calculated for the
ground-state $T=0$ K structures, and materials that undergo phase transitions around or below room temperature present particular challenges.\cite{mingo}'' Furthermore, Mingo and coworkers indicate that ``extreme materials display strong anharmonicity, and ... the inclusion of quartic and higher order anharmonicities and renormalization of harmonic phonons will perhaps be replaced by large-scale first principles molecular dynamics simulations:\cite{mingo}'' {\em this is, precisely, the road taken in this work.}


Figure \ref{fig:fig3}(a) displays the first Brillouin zone and the $k-$point sampling achieved with a 16$\times$16 supercell without interpolation. The use of this $k-$point mesh makes inclusion of Born charges unnecessary for the phonon dispersions shown here.\cite{PhysRevB.1.910}. The high-temperature phase is fourfold-symmetric,\cite{newarXiv} making the $X-$ and $Y-$points equivalent. A power spectrum at each $k-$point is obtained through a process involving two Fourier transforms and a time autocorrelation.\cite{power} To begin with, a Fourier transform of molecular dynamics velocity data into reciprocal space is performed for each of the four atoms belonging to the unit cell. These Fourier transformed velocities are subsequently time autocorrelated, and a final Fourier transform into the frequency domain provides the resonant natural frequencies $\nu_{\alpha}(\mathbf{k},T)$ demonstrated at two nearby $k-$points $\mathbf{k}$ at $T=100$ K and 230 K in Figure \ref{fig:fig3}(b).

The finite temperature phonon dispersions of a SnSe monolayer at the $k-$points along the red path in Figure \ref{fig:fig3}(a) are shown in Figure \ref{fig:fig3}(c). As seen by superimposing zero-temperature phonons drawn by red solid lines in Figure \ref{fig:fig3}(c), the power spectra data describes the vibrational frequencies, and it includes information on phonon-phonon interactions through broadening of the natural frequencies. This {\em non-perturbative} process fully incorporates anharmonicity in the phonon frequencies and phonon lifetimes, in contrast to the standard approach\cite{shengbte}---used in Ref.~\cite{FWANG15} and multiple others---whereby phonon scattering rates are determined up to third- or fourth-order in a perturbation series and added according to Matthiessen’s rule. The simplex method was used to fit central frequencies $\nu_{\alpha}(\mathbf{k},T)$ and the full width at half-max $\Delta\nu_{\alpha}(\mathbf{k},T)$ for all peaks at each $k-$point (the height $a$ indicating the relative intensity of a given natural vibrational mode at finite temperature $T$ is an additional fitting parameter) to Lorentzian functions
\begin{equation}
F_{\alpha}(\nu,\mathbf{k},T)= \frac{(\frac{\Delta\nu_{\alpha}(\mathbf{k},T)}{2})^2}{(\nu-\nu_{{\alpha}}(\mathbf{k},T))^2+(\frac{\Delta\nu_{\alpha}(\mathbf{k},T)}{2})^2}a.
\end{equation}
Additionally, a ``baseline'' background shown in red on Figure \ref{fig:fig3}(b) was fitted to a straight line to better define phonon frequency broadening. The 12 Lorentzians with largest amplitude having frequencies $\nu_{\alpha}(\mathbf{k},T)$ and linewidth $\Delta\nu_{\alpha}(\mathbf{k},T)$ provide the vibrational spectrum at each $k-$point and temperature. Phonon lifetimes are directly obtained from $\Delta\nu_{\alpha}(\mathbf{k},T)$ as: $\tau_{l,\alpha}(\mathbf{k},T) = (\pi\Delta\nu_{\alpha}(\mathbf{k},T))^{-1}$. The average value of $\tau_{l,\alpha}(\mathbf{k},T)$ was 2.6 ps across the 100 to 400 K temperature range studied here. As a result, we observe a softening of vibrational modes along the $X-M-Y$ path at frequencies between 2 and 4 THz for $T>T_c$\cite{newarXiv} in Figures \ref{fig:fig3}(b) and \ref{fig:fig3}(c). Given that the P4/nmm symmetry group is tetragonal, and in accordance with Figure \ref{fig:fig3}(a), the $Y-$point turns into $X$ for $T$ above $T_c$ in Figure \ref{fig:fig3}(c).

Importantly, and unlike the results obtained within the standard approach to thermoelectricity,\cite{shengbte} the phonon spectrum in Figure \ref{fig:fig3}(c) does reflect the effects of the structural transformation\cite{other4,newarXiv,SHIVA19,KAICHANG} showcased earlier on in Figure \ref{fig:fig1}, thus unleashing a new approach for thermoelectricity that takes finite-temperature electronic and vibrational information into heart. The novel use of finite-temperature information will modify the predictions of thermoelectric behavior\cite{FWANG15} as obtained within the standard framework.\cite{boltztrap,boltztrap2,shengbte}

\subsection{Anomalous lattice thermal conductivity}

The electronic contribution to the thermal conductivity $\kappa_{e}$ seen in Figure \ref{fig:fig2}(c) was only as large as 0.25 W\,m$^{-1}$\,K$^{-1}$ at 400 K. Obtaining the lattice thermal conductivity $\kappa_l$ requires extracting the phonon group velocity  ${v}^i_{\alpha}(\mathbf{k},T)$ (with $i=x,y$) from the finite-temperature phonon data discussed in previous subsection. ${v}_{\alpha,i}(\mathbf{k},T)$ (with $i=x,y$) is calculated by finite-differences for each band $\alpha$ at two consecutive $k-$ points $\mathbf{k}$ and $\mathbf{k}'$ for a given $T$: $v_{\alpha,i}(\mathbf{k},T)=2\pi\frac{\nu_{{\alpha}}(\mathbf{k},T)-\nu_{{\alpha}}(\mathbf{k}',T)}{|k_i-k'_i|}$. This way, the lattice thermal conductivity $\kappa_l$ displayed in Figure \ref{fig:fig3}(d) is given by:
\begin{equation}\label{eq:ltc}
\kappa_{l,ij}(T) = \frac{1}{\Omega(T)}\sum_{\mathbf{k},\alpha} v_{\alpha,i}(\mathbf{k},T) v_{\alpha,j}(\mathbf{k},T)  \tau_{l,\alpha}(\mathbf{k},T) C_{ph,\alpha}(\mathbf{k},T),
\end{equation}
with $\tau_{l,\alpha}(\mathbf{k},T)$ the phonon lifetime and $C_{ph,\alpha}(\mathbf{k},T)$ the phonon heat capacity for the $\alpha$-th mode,
\begin{equation}\label{eq:cv}
C_{ph,\alpha}(\mathbf{k},T) =
k_B\bigg(\frac{h\nu_{\alpha}(\mathbf{k},T)}{k_BT}\bigg)^2\frac{e^{h\nu_{\alpha}(\mathbf{k},T)/k_BT}}{(e^{h\nu_{\alpha}(\mathbf{k},T)/k_BT}-1)^2}.
\end{equation}
The reader should note that even though $C_{ph}$ is obtained within the harmonic approximation, the velocities and frequencies are obtained from molecular dynamics and are thus ``renormalized'' in the sense that they include anharmonic contributions by construction.

The peak shown in Figure \ref{fig:fig3}(d) near the transition temperature is reminiscent of the anomalous lattice thermal conductivity experimentally observed in SmBaMn$_2$O$_6$ single crystals across their structural transition,\cite{CHEN19} and it constitutes one of the main results of the present theoretical work. In the ferroelectric $Pnm2_1$ phase below $T_c=212$ K, $\kappa_l$ decreases with a $\propto T^{-1}$ behavior. The lattice thermal conductivity is comparable with $\kappa_E$ close to $T_c$, at which point $\kappa_l$ exhibits a sudden increase across the transition and an enhanced $\kappa_l$ for $T>T_c$ compared to the usual method. Due to the enhanced symmetry above $T_c$, we present the average lattice thermal conductivity for $T>220$ K. Above $T_c$, the fits to the data along both directions are not statistically different. A study of Cu$_2$Se, Cu$_2$S, Ag$_2$S, and Ag$_2$Se indicated a substantially reduced lattice thermal conductivity just before the onset of their temperature dependent phase transition. \cite{CHEN18} In agreement with Ref.~\cite{AGNE19}, our results actually show a larger $\kappa_l$ past $T_c$ than the usual method, which ignores the transition altogether.

Optical phonons have been found to contribute to $\kappa_l$ substantially;\cite{SHIGA12,BEECH10,TIAN11} in other monochalcogenide systems the contribution owing to the optical modes can be greater than 20\% of the total $\kappa_l$,\cite{SHUL17} and it can be as high as 30\% in bulk SnSe.\cite{QIN16} Even just on the basis of Equation \ref{eq:ltc}, the unmitigated increase in the phonon velocities, lifetimes, and softened frequencies dominates the lattice thermal conductivity despite a saturating heat capacity $C_{ph}$. An enhanced $\kappa_l$ has been attributed to higher velocity softened phonon modes both in silica \cite{ARAM17} and in double-perovskite SmBaMn$_2$O$_6$ single crystals.\cite{CHEN19}

\begin{figure}[!tb]
\begin{center}
\includegraphics[width=0.48\textwidth]{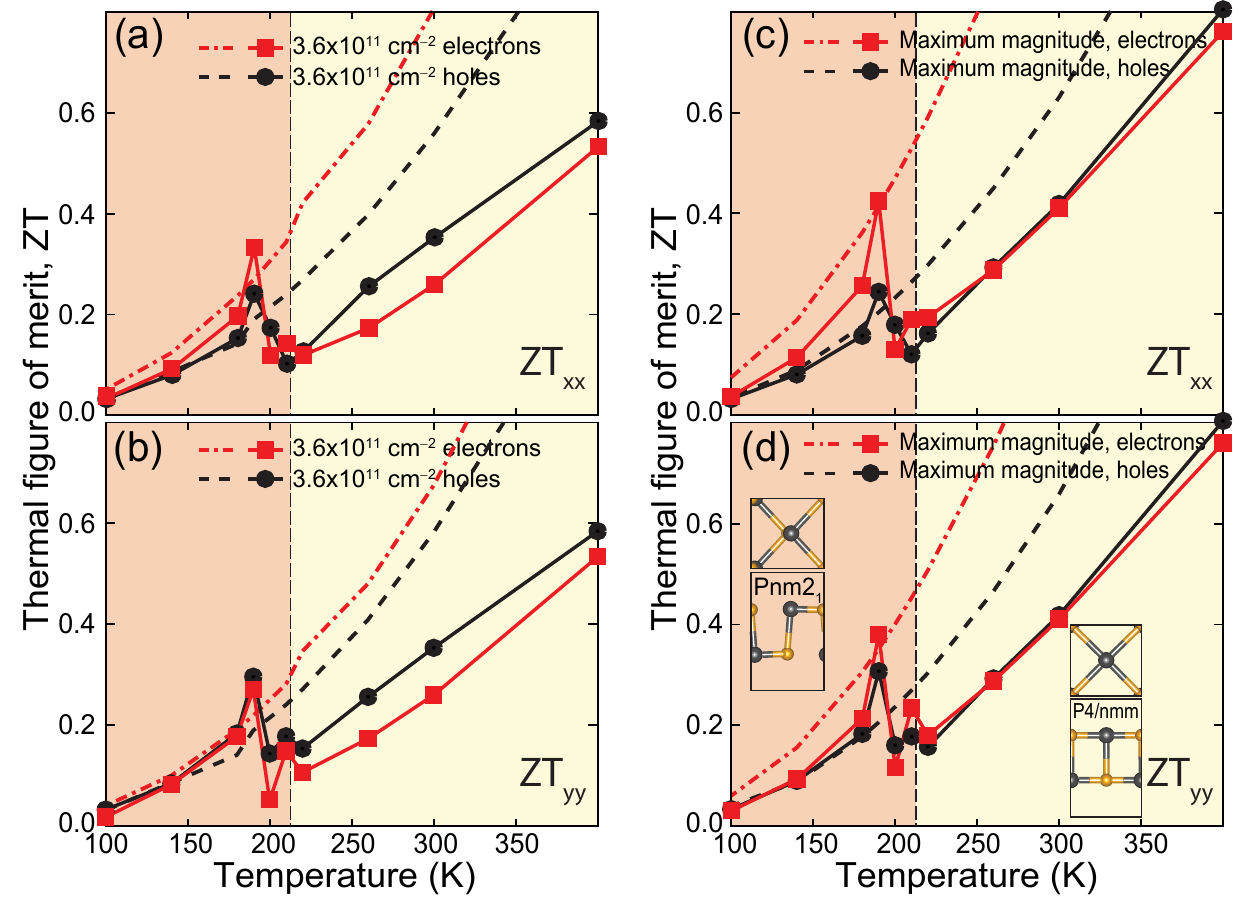}
\caption{Thermoelectric figure of merit $ZT$ for a SnSe monolayer as a function of $T$ for electron (hole) doping in red rectangles (black circles) within the method presented here. The predictions using the standard method appear smoother and are shown by dashed-dot (dashed) curves for electrons (holes). (a) $ZT$ along the $x-$direction and (b) along the  $y-$direction for a fixed carrier density of 3.6$\times 10^{11}$ cm$^{-2}$. (c) $ZT$ along the $x-$direction and (d) along the $y-$direction for a varying level of doping which maximizes $ZT$.  Note the decreased magnitude of $ZT$ past $T_c$ with respect to the standard paradigm that takes zero-temperature electronic and vibrational inputs.}\label{fig:fig4}
\end{center}
\end{figure}

\subsection{The thermoelectric figure of merit $ZT$}

Having collected $\sigma$, $\kappa_e$, $S$, and $\kappa_l$ in previous subsections, the thermoelectric figure of merit as written in Equation \ref{eq:eq1} is determined along the $x-$ and $y-$directions in Figure \ref{fig:fig4}. Subplots (a) and (b) display $ZT$ for a fixed electron or hole density along the $x-$ and $y-$directions, respectively, while subplots (c) and (d) display $ZT$ along the $x-$ and $y-$directions for electron and hole densities such that the chemical potential $\mu$ maximizes $ZT$ at each temperature.

Regardless of type of doping (electron, dashed-dotted lines or hole, dashed lines) or the amount of doping, $ZT_{xx}\ne ZT_{yy}$ for all $T$ within the standard approach to thermoelectricity. This result is due to the use of the zero-temperature rectangular ground state atomistic structure regardless of temperature. Using the finite temperature data constructed from Figure \ref{fig:fig1} to Figure \ref{fig:fig3}, the thermoelectric figure of merit $ZT$ is slightly  smaller than predictions within the zero-temperature electronic and vibrational dispersion canon just described for temperatures up to 180 K; $ZT$ grows in essentially the same manner and in agreement with the predictions of the standard paradigm. $ZT$ then displays a significant, symmetric drop beyond 190 K and for temperatures above $T_c$, which is quite different from the ultra-high magnitudes of $ZT$ reported in previous works\cite{FWANG15,SUN19,DING2019} that do not take the structural transformation into account. The apparent spike in $ZT$ near the transition temperature is similar to the behavior of $ZT$ brought on by phonon critical scattering in iodine-doped or alloyed bulk Cu$_2$Se.\cite{LIUadma13,BYEON19,VASILEV18} Though illustrated with a SnSe monolayer, the methods described here apply to any material. Their generality and relative ease of implementation make this a viable theoretical framework to understand how thermoelectric properties evolve at the onset of structural transitions.

\section{Conclusion}

We investigated the thermoelectric behavior of a prototypical SnSe monolayer across its two-dimensional ferroelectric-to-paraelectric phase transition. We incorporated finite-temperature molecular dynamics data to inform both the electronic lattice thermal behavior. As a result, $ZT$ dropped below values predicted using the standard approach to thermoelectricity past the critical temperature $T_c$. The thermoelectric behavior we have examined and the methodology we have employed are extensible to arbitrary materials undergoing solid-to-solid phase transitions. As explicitly requested by the community, the present work thus solves a pressing problem within the theory of thermoelectric materials and it provides a viable route to understand their properties when undergoing structural phase transitions.

\section{Methods}
\emph{Ab initio} molecular dynamics calculations on SnSe monolayers employing the \emph{SIESTA} code \cite{siesta} with vdW-DF-cx van der Waals corrections\cite{soler,BH}
were carried out on $16 \times 16$ supercells (originally built on a single Pnm2$_1$ phase with no domain walls). These calculations employed localized numerical atomic orbitals \cite{Junquera2001} and norm-conserving
Troullier-Martins pseudopotentials \cite{Troullier} tuned in-house \cite{rivero}, and ran with the isothermal-isobaric (NPT) ensemble for temperatures between 0 and 400 K. The out-of-plane direction of the unit cell had a length of 22 \AA~ to ensure no interaction between periodic copies of the monolayer. For comparison purposes, the ShengBTE code \cite{shengbte} was also employed to calculate the lattice thermal conductivity; the interatomic forces were calculated using the same settings employed in \emph{SIESTA} on a $5 \times 5\times 1$ supercell with up to third neighbors for the third-order force constants. We used a $36 \times 36 \times 1$ k-point mesh and a \texttt{scalebroad} parameter of 1.0.

The other factors for the thermoelectric transport coefficients are electrical. We take the average structure from each temperature in the MD calculation and use \emph{SIESTA} to
calculate the HSX file (for the Hamiltonian and overlap matrices). We use these to calculate the density of states (DOS) and necessary derivatives from a dense $k-$point mesh
of 100 points along each reciprocal lattice vector. The transport coefficient equations are given from Boltzmann transport theory.\cite{boltztrap,boltztrap2,shengbte}

Acknowledgements
The authors were funded by an Early Career Grant from the U.S.~DOE  (DE-SC0016139). Calculations were performed on Cori at NERSC, a U.S.~DOE Office of Science User Facility
(DE-AC02-05CH11231) and at the University of Arkansas' Trestles and Pinnacle supercomputers, funded by the NSF, the Arkansas Economic Development Commission, and the Office
of the Vice Provost for Research and Innovation.


%

\end{document}